\documentclass{article}
\usepackage{spconf,amsmath,amssymb,graphicx}
\usepackage{amsfonts}
\usepackage{multirow}
\usepackage{verbatim}
\usepackage{booktabs}
\usepackage{color}

\newcommand{\PTrial}{$\checkmark$}
\newcommand{\Penroll}{$\times$}

\newcommand{\Nenroll}{$\times$}

\title{Attention Back-end for Automatic Speaker Verification with Multiple Enrollment Utterances}

\name{Chang Zeng$^{1,2}$, Xin Wang$^1$, Erica Cooper$^1$, Xiaoxiao Miao$^1$, Junichi Yamagishi$^{1,2}$}
\address{$^1$National Institute of Informatics, Japan $^2$SOKENDAI, Japan}
\begin{document}
\ninept
\maketitle
\begin{abstract}
Probabilistic linear discriminant analysis (PLDA) or cosine similarity have been widely used in traditional speaker verification systems as back-end techniques to measure pairwise similarities. 
To make better use of multiple enrollment utterances, 
we propose a novel attention back-end model, which can be used for both text-independent (TI) and text-dependent (TD) speaker verification, and employ scaled-dot self-attention and feed-forward self-attention networks as architectures that learn the intra-relationships of the enrollment utterances.  
In order to verify the proposed attention back-end, we conduct a series of experiments on CNCeleb and VoxCeleb datasets by 
combining it with several sate-of-the-art speaker encoders including TDNN and ResNet. 
Experimental results using multiple enrollment utterances on CNCeleb show that the proposed attention back-end model leads to lower EER and minDCF score than the PLDA and cosine similarity counterparts for each speaker encoder 
and an experiment on VoxCeleb indicate that our model can be used even for single enrollment case. 
\end{abstract}
\section{Introduction}
\vspace{-1mm}
\label{sec:intro}

Automatic speaker verification (ASV) is a task that determines whether input speech is uttered by a claimed speaker or not. 
Until recently, generative models such as GMM-UBM \cite{Reynolds2000-GMMUBM} and i-vector \cite{Dehak2010-IV} with PLDA \cite{Ioffe2006-PLDA} were dominant for this task. With the development of deep learning, these models have been replaced by neural networks.
A good example is neural speaker embeddings \cite{Variani2014-DV, Snyder2017-DNNE} extracted from a hidden layer of a neural speaker encoder network, such as x-vectors \cite{Snyder2018-XV, Raj2019-XVProbe}, and it is reported that these have outperformed conventional i-vectors. For the case of x-vectors, speaker embedding vectors are extracted from the first fully-connected layer after a statistical pooling layer of TDNN \cite{Snyder2017-DNNE}. PLDA is adopted as a back-end model for handling the channel mismatch between enrollment speakers and evaluation speech. In addition to TDNN, ResNet \cite{He2016-ResNet} with metric learning loss \cite{Li2017-DS, defence} or discriminative training loss of softmax \cite{Liu2017-Angular, Wang2018-AM} is also a popular choice for extracting speaker embeddings. Some of them use 
simple cosine similarity between enrollment speakers and evaluation speech as a back-end model. 

The objective of this paper is to explore a new way of better handling multiple enrollment utterances and then propose a new back-end model that can use a varying number of embedding vectors corresponding to multiple enrollment utterances directly. 
In a real-world setting, enrollment speakers may provide multiple utterances to register their identity. 
We hypothesize that intra-variations of enrollment utterances contain important information such as expected acoustic variations of the enrolled speakers. Even though the traditional generative PLDA can combine multiple enrollment utterances for a better representation of a speaker, there is still room for fully using intra-variations of enrollment utterances.
Therefore, we propose a new back-end model based on scaled-dot self-attention and feed-forward self-attention networks to learn intra-variations. 

In the proposed back-end model, multiple speaker embedding vectors are first stacked as a matrix that is used as the input for multi-head scaled-dot self-attention, which estimates the self-attention probabilities of the input matrix. The self-attention matrix is then multiplied with the speaker embedding matrix in order to get bias parameters that emphasize or de-emphasize not only multiple enrollment utterances but also individual dimensions of their embedding vectors. The second step is multi-head feed-forward self-attention for aggregating the matrix modified using the multi-head scaled-dot self-attention described above into one representative vector. Finally, cosine scoring using the representative vector and one extracted from an evaluation trial is done with a learnable logistic transform for better score calibration. This network is trained using the weighted loss of normal binary cross-entropy and a recently proposed generalized end-to-end (GE2E) loss for ASV \cite{ge2e}. Note that our attention back-end model, which uses attention mechanism on the \textit{utterance-level}, is different from previous work \cite{asp,povey-sa} in which attention mechanism was applied to \textit{frame-level} aggregation.

Experiments are conducted on two ASV datasets, including the well explored VoxCeleb \cite{Nagrani2017-VOX1,Chung2018-VOX2} and new multi-genre CNCeleb \cite{Fan2020-CN1, Li2020-CN2}. Our proposed attention-based back-end is first evaluated for a single enrollment case on VoxCeleb and further analyzed for a multiple enrollment case on CNCeleb. 
The remainder of the paper is organized as follows. Cosine similarity and PLDA are briefly reviewed in Section 2. Section 3 gives details on the proposed attention back-end model. The experimental setup and results are outlined in Section 4. Finally, conclusions are presented in Section 5.
\vspace{-1mm}

\section{Conventional back-end models}
\vspace{-1mm}
\subsection{Cosine similarity}
\vspace{-1mm}

Cosine similarity can be used to evaluate how close of two speaker embeddings distributed in hyper-sphere and this serves as the simplest back-end. It has been frequently used as a scoring method for ASV systems trained using discriminative loss functions \cite{Li2017-DS, Liu2017-Angular, Wang2018-LMCL} that minimizes intra-class distance and maximizes inter-class distance. This scoring method is just an inter-product of a pair of speaker embedding vectors\footnote{In this paper, all vectors are assumed to be column vectors.} to be scored and computed as follows.
\begin{equation}
\label{cosine}
P(\boldsymbol{e}_i, \boldsymbol{e}_j)=\frac{\boldsymbol{e}_i^\top \boldsymbol{e}_j}{||\boldsymbol{e}_i||\, ||\boldsymbol{e}_j||}
\end{equation}
%
As for the case of multi-session enrollment, which means using multiple enrollment utterances per speaker, 
a frequently made choice is to average speaker embedding $\boldsymbol{c}_n = \frac{1}{K}\sum_{k=1}^K \boldsymbol{e}_m$. Here, $K$ is the number of enrollment utterances per speaker. Alternatively, a speaker embedding may be extracted from concatenated waveform or acoustic features.
\vspace{-1mm}
%

\subsection{Probabilistic linear discriminant analysis}
\vspace{-1mm}

Probabilistic Linear Discriminant Analysis (PLDA) is also a frequently used back-end model. PLDA was first introduced to ASV in \cite{first_plda} to decompose the speaker embedding of an utterance $n$ into a global mean $\boldsymbol{\mu}$, latent variable $\boldsymbol{h}_n$ and residual term $\boldsymbol{\epsilon}_{n}$ as shown below.
\begin{equation}
\label{plda}
\boldsymbol{e}_{n}=\boldsymbol{\mu}+\boldsymbol{F}\boldsymbol{h}_n+\boldsymbol{\epsilon}_{n}
\end{equation}
%
In Gaussian PLDA (GPLDA), where both the latent variable $\boldsymbol{h}$ and residual term $\boldsymbol{\epsilon}$ are assumed to follow Gaussian distributions, that is, $\boldsymbol{h} \sim \mathcal{N}(0, \mathbf{I})$ and $\boldsymbol{\epsilon} \sim \mathcal{N}(0, \mathbf{\Sigma})$, where $\mathbf{I}$ and $\boldsymbol{\Sigma}$ represent identity and covariance matrices, respectively, the score of a pair of speaker embeddings is given as follows for a single-session enrollment case:
\begin{align}
    \label{plda_s}
    P(\boldsymbol{e}_{i},\boldsymbol{e}_{j}) & = \boldsymbol{e}_{i}^\top \boldsymbol{Q}\boldsymbol{e}_{i}+\boldsymbol{e}_{j}^\top \boldsymbol{Q}\boldsymbol{e}_{j}+2\boldsymbol{e}_{i}^\top \boldsymbol{P}\boldsymbol{e}_{j} \\
    \boldsymbol{P} & = \boldsymbol{\Sigma}^{-1}_{tot}\boldsymbol{\Sigma}_{ac}(\boldsymbol{\Sigma}_{tot}-\boldsymbol{\Sigma}_{ac}\boldsymbol{\Sigma}^{-1}_{tot}\boldsymbol{\Sigma}_{ac})^{-1}  \\ 
    \boldsymbol{Q} & =\boldsymbol{\Sigma}^{-1}_{tot}-(\boldsymbol{\Sigma}_{tot}-\boldsymbol{\Sigma}_{ac}\boldsymbol{\Sigma}^{-1}_{tot}\boldsymbol{\Sigma}_{ac})^{-1} \\
    \boldsymbol{\Sigma}_{tot} & =\boldsymbol{F}\boldsymbol{F}^\top+\boldsymbol{\Sigma} \\ 
    \boldsymbol{\Sigma}_{ac} & =\boldsymbol{F}\boldsymbol{F}^\top ,
\end{align}
where $\{\boldsymbol{\mu},\boldsymbol{F},\boldsymbol{\Sigma}\}$ are trainable parameters and estimated on the basis of the EM algorithm.

There are several ways of handling the multi-session enrollment case with PLDA \cite{kong_plda, ville_plda}. According to results reported in \cite{kong_plda, ville_plda}, we adopt the averaging and concatenating of multiple speaker embeddings as well as an extended PLDA formulation \cite{ville_plda} as strategies to handle multiple enrollment utterances in our baseline model.
\vspace{-1mm}

%

\section{Attention back-end}
\vspace{-1mm}
\label{sec:pagestyle}

\begin{figure}[t]
  \centering
  \includegraphics[trim=0 80 0 70, clip, width=0.9\linewidth]{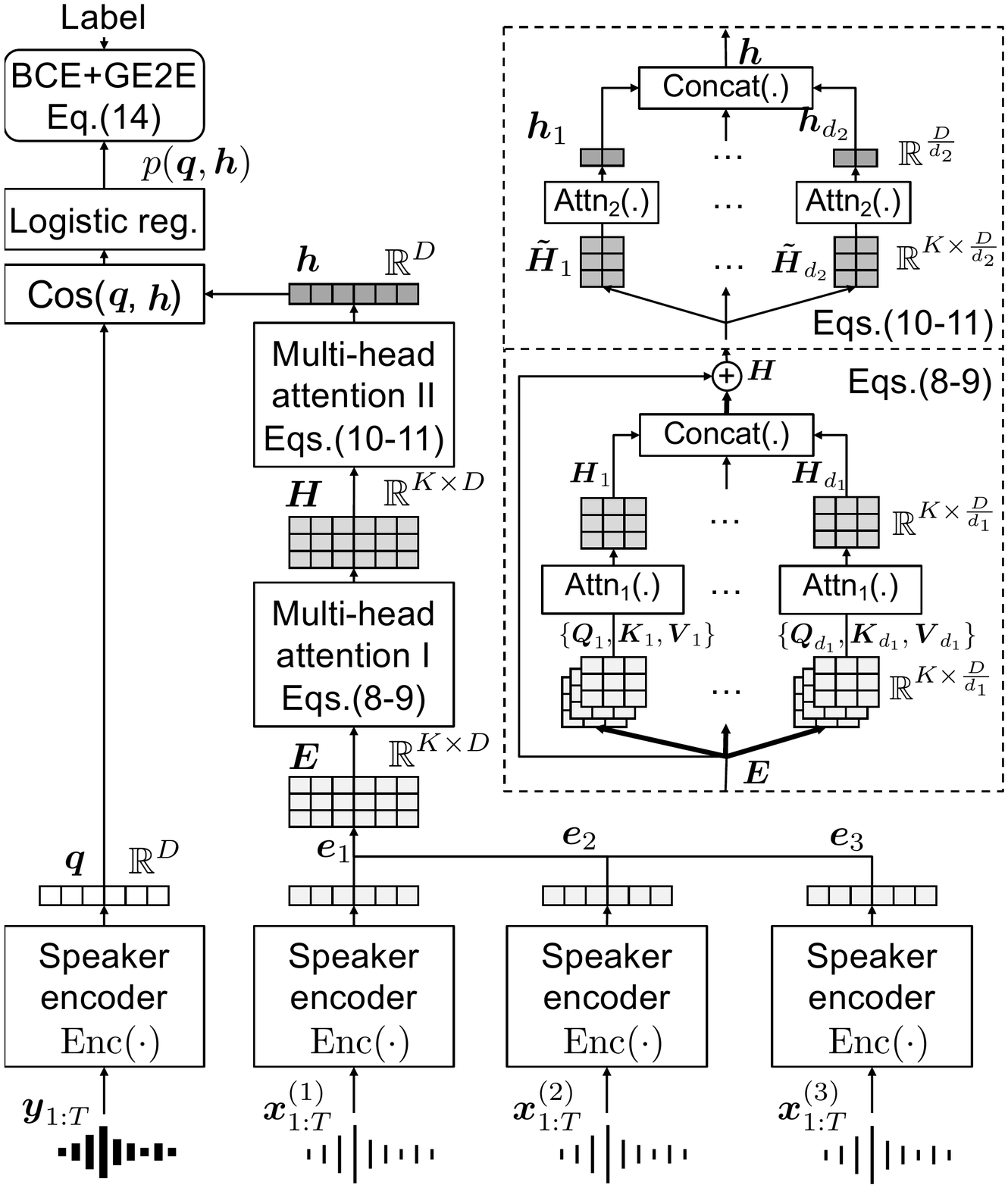}
  \caption{Back-end architecture. Dashed boxes show detailed implementation of two multi-head attention blocks.}
  \label{fig:back-end}
\end{figure}

In this section, we explain how the proposed model uses attention mechanisms to measure pairwise similarity, including the detailed architecture of the model and its loss functions, which are illustrated in Fig. \ref{fig:back-end}.
\vspace{-1mm}

\subsection{Model architecture}
\vspace{-1mm}


Suppose a speaker has $K$ enrollment utterances. With a speaker encoder $\text{Enc}(\cdot)$,  we get $K$ enrollment embedding vectors $\{\boldsymbol{e}_1, \cdots, \boldsymbol{e}_K\}$, where $\boldsymbol{e}_k=\text{Enc}(\boldsymbol{x}_{1:T}^{(k)})\in\mathbb{R}^{D}$, and $\boldsymbol{x}_{1:T}^{(k)}$ is the waveform data of the $k$-th enrollment utterance.
Since $K$ varies across speakers, we need to aggregate $\{\boldsymbol{e}_1, \cdots, \boldsymbol{e}_K\}$ into a single vector $\boldsymbol{h}\in\mathbb{R}^{D}$ using two different attention mechanisms in sequence. Specifically, after stacking a matrix $\boldsymbol{E} = [\boldsymbol{e}_1, \cdots, \boldsymbol{e}_K]^\top\in\mathbb{R}^{K\times{D}}$, we use multi-head scaled dot self-attention (SDSA) \cite{Vaswani2017-Attention} to transform $\boldsymbol{E}$ into a hidden matrix $\boldsymbol{H}\in\mathbb{R}^{K\times{D}}$. With $d_1$ attention heads, we get
\begin{equation}
    \label{multihead}
    \boldsymbol{H} = {\rm{Concat}}(\boldsymbol{H}_1, \boldsymbol{H}_2, \cdots , \boldsymbol{H}_{d_1}) \boldsymbol{W}^O + \boldsymbol{E},
\end{equation}
where the $i$-th head matrix $\boldsymbol{H}_i\in\mathbb{R}^{K\times\frac{D}{d_1}}$ is computed by
\begin{equation}
    \label{attention_transform}
    \boldsymbol{H}_i = \text{Attn}_1(\boldsymbol{Q}_i, \boldsymbol{K}_i, \boldsymbol{V}_i) = \text{Softmax}(\frac{\boldsymbol{Q}_i\boldsymbol{K}_i^\top}{\sqrt{{D}/{d_1}}})\boldsymbol{V}_i.
\end{equation}
The query, key, and value matrices are set to $\boldsymbol{Q}_i = \boldsymbol{E}\boldsymbol{W}_i^Q$, $\boldsymbol{K}_i = \boldsymbol{E}\boldsymbol{W}_i^K$, and $\boldsymbol{V}_i = \boldsymbol{E}\boldsymbol{W}_i^V$, respectively, and they all have the shape of  ${K\times\frac{D}{d_1}}$. The matrices $\{\boldsymbol{W}_i^Q, \boldsymbol{W}_i^K, \boldsymbol{W}_i^V$, $\boldsymbol{W}^O\}$ are trainable parameters of the attention mechanism. 
%
The multiple heads in Eq.\ (\ref{multihead}) are expected to attend to information from different representation subspaces at different positions.

Given $\boldsymbol{H}$ from Eq.~(\ref{multihead}), the second multi-head feed-forward self-attention (FFSA) mechanism aggregates it into the single representative embedding vector $\boldsymbol{h}$ by 
\begin{equation}
    \label{multihead_aggregation}
    \boldsymbol{h} = {\text{Concat}}(\boldsymbol{h}_1, \boldsymbol{h}_1, \cdots, \boldsymbol{h}_{d_2}).
\end{equation}
The $j$-th head vector $\boldsymbol{h}_j\in\mathbb{R}^{\frac{D}{d_2}}$ is computed as
\begin{equation}
   \label{multihead_aggregation_2}
    \boldsymbol{h}_j = \text{Attn}_2{(\boldsymbol{\tilde{H}}_j)}=\text{Softmax}(\boldsymbol{v}_j^\top \text{Tanh}(\boldsymbol{W}_j \boldsymbol{\tilde{H}}_j^\top)) \boldsymbol{\tilde{H}}_j,
\end{equation}
where $\boldsymbol{v}_j\in\mathbb{R}^{D_{2}}$ and $\boldsymbol{W}_j\in\mathbb{R}^{D_{2}\times{\frac{D}{d_2}}}$ are trainable parameters, and $\boldsymbol{\tilde{H}}_j\in\mathbb{R}^{K\times{\frac{D}{d_2}}}$ is a sub-matrix of $\boldsymbol{H}$, i.e., $\boldsymbol{H}=[\boldsymbol{\tilde{H}}_1, \boldsymbol{\tilde{H}}_2, \cdots, \boldsymbol{\tilde{H}}_{d_2}]$.
This is inspired by the multi-head attention-based aggregation proposed in \cite{Lin2017-StructureATT}.

After obtaining the representative vector for an enrollment speaker, the cosine similarity score between an evaluation speaker embedding, which is denoted as $\boldsymbol{q} = \text{Enc}(\boldsymbol{y}_{1:T})$, and the representative vector $\boldsymbol{h}$ is calculated. Further, a logistic regression (LR)  is used for better calibration as shown below:
\begin{align}
    \label{equa:logistic}
    P(\boldsymbol{q},\boldsymbol{h}) 
    &= \frac{1}{1 + \exp^{-s}} 
    = \frac{1}{1 + \exp^{-a \, \text{Cos}(\boldsymbol{q},\boldsymbol{h}) - b}} \\
    \mathrm{Cos}(\boldsymbol{q},\boldsymbol{h}) &=\frac{\boldsymbol{q} ^\top \boldsymbol{h}}{||\boldsymbol{q}||\,||\boldsymbol{h}||} ,
\end{align}
where $P(\boldsymbol{q},\boldsymbol{h})$ denotes the probability of $\boldsymbol{q}$ and $\boldsymbol{h}$ belonging to the same speaker and 
also serves as the final score for decision making, and $a$ and $b$ are trainable parameters.
\vspace{-1mm}

\subsection{Loss functions}
\vspace{-1mm}
\label{3.2}

In the training stage, speakers in a mini-batch can be seen as either target or non-target speakers, and all audio files included in one mini-batch can be used to define various positive and negative pairs.  Table~\ref{fig:train} illustrates one example of forming pairs of \textit{(test-trial, enrollment-data)}. In this example, one mini-batch of data has 3 speakers $A$, $B$, and $C$, and each speaker has 4 embeddings whose index ranges from 1 to 4. 
There are numerous ways of composing the pairs, but we here consider only the following cases. 
For positive pairs where the test trial and enrollment data are from the same speaker, one test trial is selected from the speaker's data, and the rest are left for enrollment. 
For negative pairs where the speaker of the test trial is different from that of the enrollment data, we consider only pairs marked by (test-trial=\PTrial, enroll=(\Nenroll, \Nenroll, \Nenroll)) of other speakers included in a mini-batch. 

Using the above positive and negative pairs included in a mini-batch, we optimize the proposed attention back-end. The loss function for the optimization is a weighted sum of binary cross-entropy (BCE) and generalized end-to-end (GE2E) loss \cite{ge2e} functions given below to train our model.
\begin{equation}
\label{loss}
\mathcal{L}=\lambda \mathcal{L}_{\mathrm{ge2e}} + (1-\lambda) \mathcal{L}_{\mathrm{bce}}
\end{equation}

As shown in Table~\ref{fig:train}, let us define $\boldsymbol{q}_{lm}$ as a speaker embedding vector extracted from the $m$-th test trial of speaker $l$ and $\boldsymbol{h}_{nm}$ as a vector $\boldsymbol{h}$ of Eq.\ (\ref{multihead_aggregation}) based on an enrollment set $m$, which contains multiple audio files uttered by speaker $n$. The BCE loss can be computed as  
\begin{align}
\label{bce}
\mathcal{L}_{\mathrm{bce}} = & - \sum_{^{\forall} l, m, n} \, [  \mathcal{I} \left( l=n \right) \log P ( \boldsymbol{q}_{lm},\boldsymbol{h}_{nm} ) \\ \nonumber
                           & +  \mathcal{I}\left( l \neq n \right) \log \left( 1-P \left(\boldsymbol{q}_{lm},\boldsymbol{h}_{nm}  \right) \right) ] ,
\end{align}
where $\mathcal{I}(\cdot)$ is an indicator function that returns one when its argument is true and zero otherwise. In addition to BCE, GE2E \cite{ge2e} is also used for reducing intra-class variance and enlarging inter-class variance. In the original GE2E loss, an averaged speaker embedding is computed per speaker as a centroid. Here, we treated a vector $\boldsymbol{h}$ of Eq.\ (\ref{multihead_aggregation}) as a centroid instead of a simple mean. Given this revision, the GE2E loss is given by
\begin{align}
\label{ge2e}
\mathcal{L}_{\mathrm{ge2e}} & = - \sum_{^{\forall} l, m} \log \frac{\exp^{P ( \boldsymbol{q}_{lm},\boldsymbol{h}_{lm} )}}{\sum_{^{\forall} n} \exp^{P ( \boldsymbol{q}_{lm},\boldsymbol{h}_{nm} )}}.
\end{align}
\vspace{-3mm}

\begin{table}[t]
\scriptsize
    \centering
    \caption{Composition of pairs of (test-trial, enrollment-data) for training back-end model and ground-truth labels from mini-batch. A, B, and C are speaker IDs, and 1, 2, 3 and 4 are his or her audio IDs. \PTrial and \Penroll denote test and enrollment audio files, respectively. \label{fig:train}}
    \vspace{2mm}
    \setlength{\tabcolsep}{3pt}
    {
    \begin{tabular}{ccccccccccccccccc}
 \cmidrule{2-13} 
 & \multicolumn{4}{c}{$A$} & \multicolumn{4}{c}{$B$} & \multicolumn{4}{c}{$C$} \\
 \cmidrule{2-13} 
 & 1 & 2 & 3 & 4 & 1 & 2 & 3 & 4 & 1 & 2 & 3 & 4 & & Test & Enroll & Label\\
 \cmidrule{2-13}
 \multirow{10}*{\rotatebox[origin=c]{90}{\textbf{Trials to be used for training}}} & \PTrial & \Penroll  & \Penroll & \Penroll & & & & & & & & & & $\boldsymbol{q}_{A1}$ & $\boldsymbol{h}_{A1}$ & P\\
 
 & \PTrial &  &  &  &  & \Nenroll  & \Nenroll & \Nenroll  & &  & & & & $\boldsymbol{q}_{A1}$ & $\boldsymbol{h}_{B1}$ & N \\
 
 & \PTrial &  &  &  &  & &  & & & \Nenroll  & \Nenroll & \Nenroll &  & $\boldsymbol{q}_{A1}$ & $\boldsymbol{h}_{C1}$ & N \\

 & \Penroll & \PTrial & \Penroll  & \Penroll & & & & & & & & & & $\boldsymbol{q}_{A2}$ & $\boldsymbol{h}_{A2}$ & P\\
 
 & & \PTrial &  &  & \Nenroll & & \Nenroll  & \Nenroll &  &  & & & & $\boldsymbol{q}_{A2}$ & $\boldsymbol{h}_{B2}$ & N \\
 
 & & \PTrial &  &  &  &  &  & & \Nenroll & & \Nenroll & \Nenroll & &  $\boldsymbol{q}_{A2}$ & $\boldsymbol{h}_{C2}$ & N \\

 
 
 
 & & & & &  & & \vdots & & & & &  & & \\

 & \Nenroll & \Nenroll & \Nenroll & &  &  &  &  &  &  & & \PTrial  & & $\boldsymbol{q}_{C4}$ & $\boldsymbol{h}_{A4}$ & N \\
 
 & & & &  & \Nenroll  & \Nenroll & \Nenroll &   &  &  &  & \PTrial & & $\boldsymbol{q}_{C4}$ & $\boldsymbol{h}_{B4}$ & N \\
 
 & &  &  &   & & & & &  \Penroll & \Penroll & \Penroll & \PTrial & & $\boldsymbol{q}_{C4}$ & $\boldsymbol{h}_{C4}$ & P\\

 \cmidrule{2-13}
\end{tabular}
}
\vspace{-2mm}
\end{table}

\section{Experiments}
\vspace{-1mm}
\label{sec:typestyle}

\textbf{Datasets}
We conducted our experiments on two totally different datasets: VoxCeleb1\&2 and CNCeleb1\&2, which are English and Chinese datasets, respectively. VoxCeleb1\&2 contains 7,363 speakers, whose voice comes from interviews, and is split into 7,323 speakers for training data and 40 speakers for testing data according to the official trial protocol. For trial pairs in the VoxCeleb1 test data, each speaker only has one enrollment utterance. Different from VoxCeleb, the CNCeleb1\&2 dataset includes in-the-wild speech utterances of 3,000 speakers in 11 different genres, which leads to complex and composite inter-session variations, both intrinsic (i.e., speaking style, physiological status) and extrinsic (i.e., recording device, background noise) \cite{Li2020-CN2}.  The official protocol uses 2,800 speakers for training and 200 speakers whose identity was enrolled in multiple sessions for evaluation. MUSAN \cite{musan} and RIRs \cite{rirs} datasets were both used to augment the above datasets.

\noindent
\textbf{Baseline}
Our experiments used a few well-testified DNN models to extract fixed-length speaker embeddings: TDNN \cite{Snyder2018-XV}, TDNN with attentive statistics pooling (TDNN-ASP) \cite{asp}, ECAPA-TDNN \cite{ecapa}, and ResNet \cite{He2016-ResNet}. All TDNN variants used plain softmax loss, while the ResNet used  am-softmax loss \cite{Wang2018-LMCL}. The TDNN models were trained by using an SGD optimizer with 0.01 learning rate. As for ResNet, it was trained by using an SGD optimizer with 0.1 learning rate. Both of them were trained for 20 epochs with 0.1 learning rate decay every 8 epochs. 
During the back-end training, all the speaker encoders were frozen.
The GPLDA back-end was trained by using a subset of training data, which consisted of 100,000 speaker embeddings selected randomly from the training data. First, a linear discriminative analysis (LDA) transformation matrix was trained to reduce the dimension of the speaker embedding from 512 to 400. Then, the output of LDA was used to train a GPLDA model in which the dimension of the latent factor is 150. 

\noindent
\textbf{Back-end training}
A balanced batch strategy was adopted to train the attention back-end model. For each mini-batch, assuming it has $M$ speakers and $K$ speaker embeddings per speaker, the size of one mini-batch was $M\times K$. 
In our experiment, $M$ and $K$ were set to 256 and 5, respectively. When one mini-batch of data was fed into the back-end model, it was rearranged to form speaker verification trials that had multiple enrollment utterances (or embedding vectors) as described in Section \ref{3.2} and the ratio of positive samples to negative samples was $1:(M-1)$. 
The proposed model was trained for 40 epochs by using an SGD optimizer with a cyclical learning rate scheduler \cite{cyclic}, which changed learning rate from 0.00001 to 0.00003 every 2,000 updating steps. The hyper-parameter $\lambda$, which is the weight of two loss functions, was set to 0.6. 
In the testing phase, the output of Eq.\ (\ref{equa:logistic}) served as a decision score.
\vspace{-1mm}


\subsection{Results}
\vspace{-1mm}

\begin{table}[t]
\vspace{-2mm}
\footnotesize
  \caption{VoxCeleb verification performance results with TDNN and different back-ends}
  \label{tab:result_vox}
  \vspace{1mm}
  \centering
  \begin{tabular}{lcccc}
    \toprule
    \textbf{Back-end} & \textbf{EER(\%)} & \textbf{minDCF(0.01)} & \textbf{minDCF(0.001)} \\
    \midrule
    Cosine     & 10.51              & 0.7928                & 0.8718                    \\
    PLDA       & \textbf{3.14}      & 0.3456                & 0.5567                    \\
    Proposed   & 3.26               & \textbf{0.3323}       & \textbf{0.5134}           \\
    \bottomrule
  \end{tabular}
  \vspace{-5mm}
\end{table}

Although the main advantage of the proposed method is that multiple enrollment utterances can be used, we first confirmed the performance of the proposed back-end using a single enrollment case on the VoxCeleb dataset. This ensures the generalization capability of our proposed back-end with respect to the number of enrollment utterances. Table \ref{tab:result_vox} shows the result for the TDNN speaker encoder on the VoxCeleb dataset. We can see that the attention back-end gave a number comparable to that of PLDA in terms of EER. Moreover, in terms of minDCF(0.01 and 0.001), the proposed model slightly outperformed PLDA model by 3.8\% and 7.8\%, respectively, in relative. Although the proposed architecture is motivated by the use of multiple enrollment utterances, it also works reasonably well for the single enrollment case. 

\begin{figure}
\vspace{-2mm}
    \centering
    \caption{DET curves for combinations of TDNN and ResNet speaker encoders and several different back-end models.}
    \label{fig:det_tdnn}
    \vspace{2mm}
    \includegraphics[width=1\linewidth]{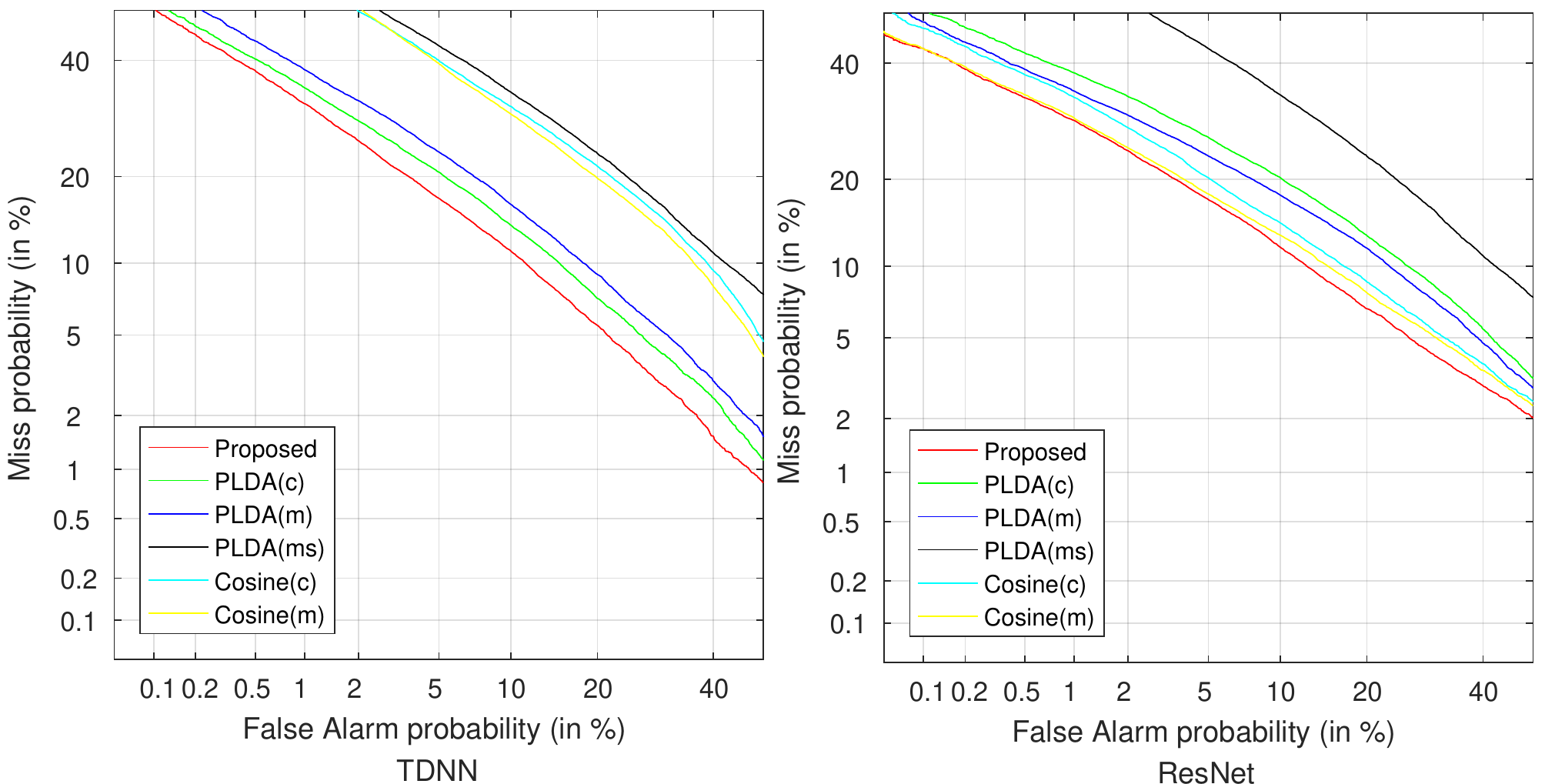}
    \vspace{-9mm}
\end{figure}

In the following experiments, we focus on CNCeleb since this dataset has multiple enrollment utterances per speaker, and the acoustic conditions of the enrollment data may vary significantly. Therefore, it is a suitable dataset for evaluating the proposed methods. The first and the second groups of Table \ref{tab:result_cn} shows the EER and minDCF(0.01) results for various back-end models using TDNN or ResNet encoders trained on CNCeleb, respectively. The third line of this table is the best result of others' work that we are aware of so far for this dataset.
From the table, we first see that PLDA variants were better than cosine similarity variants for the TDNN encoder, whereas cosine similarity variants were more suitable for the discriminatively trained ResNet encoder, as reported in the literature. We can then see that our attention back-end model had a lower EER and minDCF score than the PLDA and cosine similarity variants for each speaker encoder type. 
The DET curves of these two speaker encoders with different back-end methods are shown in Figure \ref{fig:det_tdnn}.
%

We also combined our attention back-end with TDNN-ASP and ECAPA-TDNN models, which contained frame-level attention module. The result for the third group of Table \ref{tab:result_cn} indicates that the utterance-level attention module included in our proposed model was complementary to the frame-level attention used in the speaker encoders in the third group. Also, we can confirm that our attention back-end had better performance than PLDA back-end again.
\vspace{-1mm}


\setlength{\tabcolsep}{1.0mm}
\begin{table}[t]
\footnotesize
  \caption{CNCeleb verification performance results of two encoders with different back-ends}
  
  \label{tab:result_cn}
  \vspace{2mm}
  \centering
  \begin{tabular}{l c c c c}
    \toprule
    \multicolumn{1}{c}{\textbf{Spk encoder}} & \multicolumn{1}{c}{\textbf{Back-end}} & \multicolumn{1}{c}{\textbf{Enroll process.}} & \multicolumn{1}{c}{\textbf{EER(\%)}} & \multicolumn{1}{c}{\textbf{minDCF(0.01)}}\\
    \midrule
    TDNN   & Cosine         & Concat         & 20.82           & 0.8114                \\
    TDNN   & Cosine         & Mean           & 19.86           & 0.8350                \\
    TDNN \cite{Li2020-CN2}  & PLDA & Concat  & 12.52           & -                     \\
    TDNN   & PLDA           & Concat         & 12.09           & 0.6105                \\
    TDNN   & PLDA           & Mean           & 13.29           & 0.6522                \\
    TDNN   & PLDA           & Multi-session  & 21.35           & 0.8417                \\
    TDNN   & Proposed       & Attention      & \textbf{10.12}  & \textbf{0.5649}       \\ \midrule
    ResNet & Cosine         & Concat         & 12.46           & 0.5484                \\
    ResNet & Cosine         & Mean           & 11.86           & 0.5045                \\
    ResNet & PLDA           & Concat         & 15.66           & 0.5890                \\
    ResNet & PLDA           & Mean           & 14.56           & 0.5692                \\
    ResNet & PLDA           & Multi-session  & 21.91           & 0.8082                \\
    ResNet & Proposed       & Attention      & \textbf{10.77}  & \textbf{0.4983}       \\ \midrule
    TDNN-ASP & PLDA         & Concat         & 11.63           & 0.6157                \\
    TDNN-ASP & PLDA         & Mean           & 10.67           & 0.5990                \\
    TDNN-ASP & Proposed     & Attention      & \textbf{9.90}   & \textbf{0.5606}       \\
    ECAPA-TDNN & PLDA       & Concat         & 10.54           & 0.5806                \\
    ECAPA-TDNN & PLDA       & Mean           & 9.51            & 0.5665                \\
    ECAPA-TDNN & Proposed   & Attention      & \textbf{8.93}   & \textbf{0.5043}       \\
    \bottomrule
  \end{tabular}
  \vspace{-5mm}
\end{table}

\setlength{\tabcolsep}{0.5mm}
\begin{table}[t]
\footnotesize
  \caption{Ablation experiment results on CNCeleb with TDNN speaker encoder. }
  \label{tab:ablation}
  \centering
  \vspace{1mm}
  \begin{tabular}{c c c c c}
    \toprule
    \multicolumn{1}{c}{\textbf{Systems}} & \multicolumn{1}{c}{\textbf{EER(\%)}} & \multicolumn{1}{c}{\textbf{DCF(0.01)}} & \multicolumn{1}{c}{\textbf{DCF(0.001)}}\\
    \midrule
    SDSA + FFSA           + LR + BCE + GE2E       & 10.12          & 0.5649          & 0.7127          \\
    \phantom{SDSA +} FFSA + LR + BCE + GE2E       & 19.55          & 0.8095          & 0.9046          \\
    SDSA + mean           + LR + BCE + GE2E       & 10.26          & 0.5876          & 0.7270          \\
    SDSA + FFSA \phantom{+ LR} + BCE + GE2E       & 48.32          & 1.0000          & 1.0000          \\
    SDSA + FFSA + LR \phantom{+ BCE} + GE2E       & 12.74          & 0.5937          & 0.7350          \\
    SDSA + FFSA + LR + BCE \phantom{+ GE2E}       & 10.28          & 0.5896          & 0.7402          \\
    \bottomrule
  \end{tabular}
   \vspace{-5mm}
\end{table}

\subsection{Ablation study}
\vspace{-1mm}

To explore the contribution of each component of our back-end for performance improvement, we conducted an ablation experiment whose result is shown in Table \ref{tab:ablation}. This study also used the CNCeleb dataset. The proposed model can be divided into five components (SDSA, FFSA, LR, BCE and GE2E) as shown in the first line of the table. In each ablation study, we removed one component from the model except the FFSA, which was substituted with a simple mean operation for aggregation. 

From the second and the third lines of the table, it was confirmed that two self-attention mechanisms both contributed to the improvement. Concretely, both SDSA for summarizing intra-variations of multiple speaker embeddings and FFSA for aggregation contributed to the performance, and SDSA seemed to have a more important role. Additionally, from the fourth line, we can also see that LR for calibration was another essential component of the proposed model. 
%
The first, fifth and sixth lines of the table indicates interesting tendencies of our loss function using the weighted sum of BCE and GE2E. From a comparison of the fifth and sixth lines, we see that the proposed model using BCE loss only had a lower EER than that using GE2E loss only, while one using GE2E loss resulted in a lower minDCF(0.001) value. Combining BCE and GE2E thus resulted in the lowest values for all the metrics. 
\vspace{-1mm}

\section{Conclusion}
\vspace{-1mm}
\label{sec:majhead}

In this paper, we proposed a new attention back-end model for speaker verification using a varying number of multiple enrollment utterances and compared it with other back-end methods. The proposed back-end uses two types of attention, SDSA and FFSA, 
and they are trained based on weighted BCE and GE2E loss. For the CNCeleb dataset, the proposed attention back-end model was combined with several speaker encoders including TDNN and ResNet, and they were compared with PLDA or cosine similarity counterparts. The proposed model showed superiority and resulted in a lower EER and minDCF score than the PLDA and cosine similarity methods for each speaker encoder type. 
The experimental results on the VoxCeleb dataset show that the proposed back-end has comparable performance to conventional PLDA scoring even for the case of a single enrollment utterance.

\vspace{1mm}
\noindent
\textbf{Acknowledgement}
This study is supported by JST CREST Grants (JPMJCR18A6, JPMJCR20D3), MEXT KAKENHI Grants (16H06-302, 18H04120, 18H04112, 18KT0051), Japan, and Google AI for Japan program.



\bibliographystyle{IEEEbib}
\bibliography{strings, refs}

\end{document}